# Superconductivity and Electronic Structures of Nickelate Thin Film Superstructures

Zihao Nie (聂子豪)[1†], Yueying Li (李月莹)[1,2†], Wei Lv (吕威)[1†], Lizhi Xu (徐立智)[1†], Zhicheng Jiang (江志诚)[3†], Peng Fu (付鹏)[1], Guangdi Zhou (周广迪)[1,2], Wenhua Song (宋文华)[1], Yaqi Chen (陈亚奇)[1], Heng Wang (汪恒)[1,2], Haoliang Huang (黄浩亮)[1,2], Junhao Lin (林君浩)[1,2], Jin-Feng Jia (贾金锋)[1,2,4], Dawei Shen (沈大伟)[3*], Peng Li (李鹏)[1,2*], Qi-Kun Xue (薛其坤)[1,2,5 *], Zhuoyu Chen (陈卓昱)[1,2*]

[1]State Key Laboratory of Quantum Functional Materials, Department of Physics, Guangdong Basic Research Center of Excellence for Quantum Science, and College of Semiconductors (National Graduate College for Engineers), Southern University of Science and Technology, Shenzhen 518055, China
[2]Quantum Science Center of Guangdong-Hong Kong-Macao Greater Bay Area, Shenzhen 518045, China
[3]National Synchrotron Radiation Laboratory and School of Nuclear Science and Technology, University of Science and Technology of China, Hefei 230026, China
[4]State Key Laboratory of Micronano Engineering Science, Tsung-Dao Lee Institute & School of Physics and Astronomy, Key Laboratory of Artificial Structures and Quantum Control (Ministry of Education), Shanghai Jiao Tong University, Shanghai 200240, China
[5]Department of Physics, Tsinghua University, Beijing 100084, China

[†]These authors contributed equally.

*E-mail: chenzhuoyu@sustech.edu.cn, xueqk@sustech.edu.cn, lipeng@quantumsc.cn, dwshen@ustc.edu.cn

**Abstract**

Ruddlesden–Popper (RP) nickelates have emerged as a crucial platform for exploring the mechanisms of high-temperature superconductivity[1-7]. However, the Fermi surface topology required for superconductivity remains elusive. Here, beyond the superconducting pure bilayer (2222) phase, we report the thin film growth and ambient-pressure superconductivity of monolayer-bilayer (1212) and bilayer-trilayer (2323) superstructures, together with the absence of superconductivity in monolayer-trilayer (1313) superstructure, under identical compressive epitaxial strain. The onset superconducting transition temperatures range from 46 to 50 K, exceeding the McMillan limit. Angle-resolved photoemission spectroscopy reveals key Fermi surface differences in these atomically-engineered structures. In superconducting 1212 and 2222 films, a dispersive hole-like band ($\gamma^{II}$) forms an underlying Fermi pocket, surrounding the Brillouin zone corner. In contrast, the top of the flat band ($\gamma^{III}$) is observed ~70 meV below $E_F$ in the non-superconducting 1313 films. Particularly, the superconducting 2323 films host both $\gamma^{II}$ and $\gamma^{III}$ bands. The polarization dependence of the $\gamma$ bands reveals their Ni $d_{z^2}$ origin. Our findings expand the family of ambient-pressure nickelate superconductors and establish a connection between structural configuration, electronic structure, and the emergence of superconductivity in nickelates.

## Main

The nickelate superconductors can be grouped into two distinct types. The square-planar type, represented by the infinite layer[8] and quintuple layer[9,10], features their structural and electronic resemblance to cuprate superconductors, with the similar $3d^9$ electronic configuration and $d_{x^2-y^2}$-dominated Fermi surface topology[11,12]. In contrast, the Ruddlesden–Popper (RP) nickelate superconductors with the bilayer[1,2,4,13], the trilayer[3,14,15], and the hybrid monolayer-bilayer[7,16] unit structures, are generally considered to be governed by both Ni $d_{x^2-y^2}$ and $d_{z^2}$ orbitals[1,11,12,17,18]. The participation of the $d_{z^2}$ orbital introduces a new degree of freedom absent in single-band cuprates, yet its exact role for superconductivity is debated[19-31].

The ambient-pressure superconductivity achieved in RP bilayer nickelate thin films under biaxial compressive epitaxial strain[5,6,32,33] facilitates the use of angle-resolved photoemission spectroscopy (ARPES), providing unprecedented opportunity to understand the correlation between band structures and the emergence of superconductivity[34-37]. The exceptional thermodynamic tolerance of the gigantic-oxidative atomic-layer-by-layer epitaxy (GAE) method enables the systematic control over artificial hybrid superstructures essential for comparative studies[38]. In this work, with the discovery of ambient-pressure superconductivity in epitaxial thin films with the monolayer-bilayer (1212, $(La,Pr)_5Ni_3O_{11}$) and bilayer-trilayer (2323, $(La,Pr)_7Ni_5O_{17}$) superstructures, we systematically compare the low-temperature transport properties, diamagnetism, structural characterizations, and ARPES spectra among four different films under identical compressive epitaxial strain, namely the 1212, the pure bilayer (2222), the monolayer-trilayer (1313), and the 2323 structures. The 1313 structure has a different structural configuration but the same chemical composition ($(La,Pr)_3Ni_2O_7$) as the 2222 (ref.[39]).

## Comparison of various superstructures

Figure 1 summarizes the systematic comparative study. Using the GAE method[38], the four distinct structures, i.e., 1212, 2222, 1313, and 2323, were grown on identical $SrLaAlO_4$ substrates without post-annealing common in previous studies[5,6,32,33] (see Methods and Extended Data Figs. 1 and 2 for details of sample synthesis). For the rare earth site, a La:Pr ratio of 2:1 was selected for all films, since Pr substitution effectively suppresses oxygen vacancies[40] and enhances superconducting performance[4,32]. This systematic approach reveals a structural dependence in their transport properties. Superconductivity is observed in 1212, 2222 and 2323 structures, confirmed by both zero resistivity and Meissner diamagnetic effect by mutual inductance measurements (see also Extended Data Figs. 3 and 4). The resistive onset transition temperatures ($T_c^{onset}$) of 1212, 2222 and 2323 structure are ~50 K, ~50 K, and ~46 K, respectively, all exceeding the McMillan limit. The resistivities of the 1212, 2222, and 2323 structures drop below the noise level at 30 K, 25 K, and 3.5 K, respectively. Note that a two-step superconducting transition is observed in 2323 superstructures, with the second transition occurring at ~18 K. Such a two-step feature is always seen with the second transition occurs around 10 to 20 K (Extended Data Fig. 3). Meanwhile, as shown in Extended Data Fig. 5, the temperature-dependent upper critical field and angle-dependent transport in the 2323 superstructure exhibit apparent deviation from the scenario observed in the 1212 structure (which resembles the 2222 case, see ref.[5,6,32,33]), implying the coexistence of two superconducting components. A plausible interpretation is that the trilayer spacers disrupt the global phase coherence of the bilayer blocks, thus the realization of a zero-resistance state relies on coupling the bilayer blocks through proximitized superconductivity within the trilayer units. For all these superconducting structures, ARPES measurements reveal that their corresponding Fermi surfaces are qualitatively similar (Figs. 1d, 1h, and 1p, detailed results to be discussed in Figs. 3 and 4). All of them contain an electron-like α pocket around the Brillouin zone center ($\bar{\Gamma}$), a hole-like β pocket around the zone corner ($\bar{M}$), and crucially, a prominent hole-like γ pocket surrounding the $\bar{M}$ point. In contrast, the 1313 films do not show superconductivity, instead displaying a metallic behavior with a slight upturn in resistivity at low temperatures (Fig. 1j). This absence of superconductivity is concurrent with a distinct Fermi surface topology, which consists of only the α and β pockets and misses the γ pocket (Fig. 1l). Considering the bulk-form 1313 single crystal is reported superconducting under high pressure[41,42], a larger epitaxial strain (or pressure) or cation substitution may be required to achieve superconductivity in the 1313 films.

**Structural characterizations**

Structural characterizations of the four different structures are shown in Fig. 2. Scanning transmission electron microscopy (STEM) reveals the atomic structures of the 1212, 2222, 1313 and 2323 films grown on $SrLaAlO_4$ substrates. In the high-angle annular dark-field (HAADF) images (Figs. 2a, 2d, 2g, and 2j), distinct alternating sequences are observed: bilayer/monolayer blocks in the 1212 phase, bilayer/bilayer blocks in the 2222 phase, trilayer/monolayer blocks in the 1313 phase, and bilayer/trilayer blocks in the 2323 phase. The pure-phase HAADF images with a larger field of view are shown in Extended Data Fig. 6. The structural distinctions are further supported by atomically resolved energy-dispersive X-ray spectroscopy (EDS), showing element-sensitive stacking distributions (Figs. 2b, 2e, 2h and 2k). X-ray diffraction (XRD) analysis confirms that all films are single-phase and highly crystalline, exhibiting sharp diffraction peaks with no evidence of secondary phases at the macroscopic level (Figs. 2c, 2f, 2i and 2l). The presence of Kiessig fringes around the diffraction peaks further indicates smooth surfaces and sharp interfaces. For the superconducting 1212 films grown on $SrLaAlO_4$ substrates, the XRD out-of-plane lattice constant is calculated to be 16.78 Å, experiencing an ~1% elongation compared with the bulk value of 16.575 Å (ref.[7]). The STEM-HAADF images reveal coherent in-plane strain throughout the films, resulting in a compressive strain of ~2% relative to the bulk, a value identical to that of the bilayer films. The XRD out-of-plane lattice constants for 2222, 1313 and 2323 are 20.75 Å, 20.57 Å, and 24.63 Å, respectively.

Atomically precise nickelate superstructures are ideal for comparative ARPES studies, but probing their intrinsic electronic structure is exceptionally challenging. This is rooted in the material's extreme sensitivity to oxygen stoichiometry[5,6,32,33]: oxygen deficiency is known to induce a superconductor-insulator transition, reverse the sign of the Hall coefficient[43], and qualitatively alter X-ray absorption spectra[44]. Since any oxygen loss inevitably begins at the surface, a drastically different electronic state is expected to form there first, posing risks for surface-sensitive probes like ARPES. Such a modified surface layer would be invisible to transport measurements that reflect the underlying superconducting film, creating a potential for misinterpretation. To overcome this, all films for ARPES measurements were grown on the same day under identical conditions, and transferred from the growth chamber to the synchrotron using a cryogenic ultrahigh-vacuum suitcase. This procedure involves rapid quenching of the films from their growth temperature to below 200 K immediately after synthesis, effectively "freezing" the oxygen content on the surface. The samples are then maintained below 200 K throughout the entire transfer and measurement process. This meticulous cryogenic protocol ensures the intrinsic electronic properties of the superconducting surface can be reliably probed.

**Electronic band structures**

Figure 3 displays the ARPES measurements of the four engineered nickelate thin films. To mitigate matrix element effects and to comprehensively map the Fermi surface, measurements were conducted across varied photon energies. The Fermi surface maps (Figs. 3a and 3c) reveal that the superconducting 1212, 2222 films possess Fermi surfaces composed of multiple pockets. Besides the α and β pockets, there is another prominent pocket γ centered at the zone corner $\bar{\mathrm{M}}$ point. More information is provided in Figs. 3b and 3d showing the spectral cuts along high-symmetry directions $\bar{\mathrm{M}}$-$\bar{\mathrm{X}}$-$\bar{\mathrm{M}}$, overlaid with the fitted peak positions (see Extended Data Figs. 7 and 8 for MDCs, EDCs, and the corresponding fitting methods). The spectral intensity of the β band is more pronounced at a photon energy of 103 eV, whereas the γ band shows enhanced intensity at 153 eV. This significant intensity variation cannot be simply attributed to $k_z$ dispersion, but likely arises from the distinct photoemission cross-sections of different orbitals with varying photon energies. For the superconducting 1212 and 2222 films, the $\gamma$ band forms an underlying Fermi pocket, consistent with our previous observation in the superconducting 2222 sample[34]. The 1212 superconductor has been studied theoretically[16,45-48]. Interestingly, an additional vertical feature near the $\overline{\mathrm{M}}$ point is prominent (yellow arrows). The possible origin of this feature is to be revealed through analysis of the ARPES spectra from the 1313 structure.

The ARPES spectra for the non-superconducting 1313 structure show distinct behavior from the superconducting ones. The $\alpha$ and $\beta$ bands in Figs. 3e and 3f are qualitatively similar to the 1212 and 2222 films. However, the $\gamma$ band measured at 153 eV in 1313 phase exhibits clear flat-band feature with band top ~70 meV below $E_F$, which resembles the non-superconducting 1313 bulks[49]. The spectral intensity around the $\overline{\mathrm{M}}$ point in the Fermi surface map of 1313 phase (Fig. 3e right panel) originates from the broadening of the $\gamma$ band top.

A prominent, nearly vertical feature with strong spectral intensity is also observed near the Brillouin zone corner $\overline{\mathrm{M}}$ point in the 1313 structure, extending toward higher binding energies (yellow arrows, right panels of Figs. 3f). This feature is likely a waterfall-like phenomenon, appearing as a momentum-independent intensity tail cascading from the small electron-like dip at the $\gamma$ flat band top near the $\overline{\mathrm{M}}$ point (visible in Fig. 3f). This dip is predicted in calculations considering epitaxial strain effect[31,50], and its evolution into a waterfall-like tail may be attributed to strong correlations in the $d_{z^2}$ orbital. In the superconducting 1212, 2222 cases (Figs. 3b and 3d), the top of the $\gamma$ flat band rises above $E_F$, rendering this waterfall-like tail only emerging near or below $E_F$. This finding is only made apparent through systematic comparison among different structural variants, and would be difficult to discern from 2222 data in isolation. The waterfall-like behavior is also apparent in the $\beta$ bands as they disperse toward the $\overline{\mathrm{X}}$ point in Figs. 3b, 3d, 3f, and 3h. Analogous waterfall-like features have also been reported in both infinite-layer nickelates and cuprates.

In the superconducting 2323 superstructure, although the Fermi surface topology is qualitatively similar to the superconducting 1212 and 2222 (Fig. 3g), intriguingly, two separated $\gamma$ bands are resolved in Fig. 3h: one $\gamma^{II}$ forming underlying Fermi pocket (similar to 1212 and 2222), and another $\gamma^{III}$ lying below $E_F$ (similar to 1313), together with the intense vertical waterfall-like features. Given that the monolayer structure is intrinsically insulating and not contributing directly to the photoemission intensity near $E_F$, a plausible interpretation is that bilayer and trilayer structural blocks host $\gamma^{II}$ and $\gamma^{III}$ bands, respectively, under the current straining and oxidation condition.

To further confirm the structural and orbital origin of the $\gamma^{II}$ and $\gamma^{III}$ bands, polarization-dependent ARPES measurements were performed on the superconducting samples at 103 eV along $\bar{M}$-$\bar{\Gamma}$-$\bar{M}$ in 2323 structure, shown in Fig. 4 (see Extended Data Fig. 9 for MDCs and EDCs). Measurement geometry matrix element effect analysis indicates that linear vertical (LV) polarization enhances $d_{z^2}$ orbital intensity, as compared to linear horizontal (LH) polarization (Extended Data Fig. 10). The polarization-dependent intensity variations of the α/β and γ bands confirm their $d_{x^2-y^2}$ and $d_{z^2}$ orbital association, respectively. Under the LV polarization, the α/β band still exhibits residual intensity, likely due to the contribution from hybridized oxygen *p* orbitals. The coexistence of both $\gamma^{II}$ and $\gamma^{III}$ bands is clearly resolved in Figure 4b with weaker waterfall-like feature than Figure 3h, highlighting the potential structural assignments of the $\gamma^{II}$ and $\gamma^{III}$ features to the bilayer and trilayer blocks, respectively. The polarization-dependent measurements performed on the 1212 and 2222 structures exhibit similar results without the flat $\gamma^{III}$ band (Extended Data Fig. 10). These systematic comparative results point to the relevance of the position of $d_{z^2}$ orbital for the emergence of the superconductivity, under the current strain and oxidation conditions (Table 1).

**Discussion**

Probing the intrinsic electronic structure of superconducting RP nickelate films with ARPES requires stringent control over two decisive factors. First, superconductivity is highly sensitive to oxygen stoichiometry. Oxygen defects at different sites may impact the electronic structures in distinct ways: planar vacancies disrupt in-plane phase coherence, particularly for $d_{x^2-y^2}$-derived bands; apical vacancies affect the $d_{z^2}$-derived band; and interstitial oxygen induces hole doping[40]. Our protocol, with samples fully oxidized during growth and preserved via *in situ* cryogenic ultrahigh-vacuum transfer, kinetically freezes the oxygen configuration at the surface and ensures the access to intrinsic properties by surface-sensitive ARPES. Second, the RP bilayer phase is thermodynamically less stable than the monolayer or trilayer, making it prone to the formation of secondary phases, most notably the trilayer phase. While surface impurities do not compromise underlying superconductivity across the film thickness, they are detrimental to surface-sensitive ARPES measurements. By comparing 1212, 2222, 1313, and 2323 films, we observe distinct band dispersions, which self-consistently confirms that our results capture the intrinsic electronic structures of pure structural phases.

In summary, we demonstrate ambient-pressure superconductivity in 1212 and 2323 nickelate superstructures, evidenced by both zero resistance and Meissner diamagnetic effect, with onset transition temperatures exceeding the McMillan limit. A systematic comparison across 1212, 2222, 1313, and 2323 structures identifies the position of the $d_{z^2}$-derived γ band relative to the Fermi level as the critical electronic distinction between superconducting and non-superconducting variants. These findings not only expand the family of nickelate superconductors but also establish an atomically precise platform to systematically investigate the electronic genes governing high-temperature superconductivity.

## References

1 Sun, H. *et al.* Signatures of superconductivity near 80 K in a nickelate under high pressure. *Nature* **621**, 493-498 (2023).

2 Zhang, Y. *et al.* High-temperature superconductivity with zero resistance and strange-metal behaviour in $La_3Ni_2O_{7-\delta}$. *Nat. Phys.* **20**, 1269-1273 (2024).

3 Zhu, Y. *et al.* Superconductivity in pressurized trilayer $La_4Ni_3O_{10-\delta}$ single crystals. *Nature* **631**, 531-536 (2024).

4 Wang, N. *et al.* Bulk high-temperature superconductivity in pressurized tetragonal $La_2PrNi_2O_7$. *Nature* **634**, 579-584 (2024).

5 Ko, E. K. *et al.* Signatures of ambient pressure superconductivity in thin film $La_3Ni_2O_7$. *Nature* **638**, 935-940 (2025).

6 Zhou, G. *et al.* Ambient-pressure superconductivity onset above 40 K in $(La,Pr)_3Ni_2O_7$ films. *Nature* **640**, 641-646 (2025).

7 Shi, M. *et al.* Pressure induced superconductivity in hybrid Ruddlesden–Popper $La_5Ni_3O_{11}$ single crystals. *Nat. Phys.* **21**, 1780-1786 (2025).

8 Li, D. *et al.* Superconductivity in an infinite-layer nickelate. *Nature* **572**, 624-627 (2019).

9 Pan, G. A. *et al.* Superconductivity in a quintuple-layer square-planar nickelate. *Nat. Mater.* **21**, 160-164 (2021).

10 Yan, X. *et al.* Superconductivity in an ultrathin multilayer nickelate. *Sci. Adv.* **11**, eado4572 (2025).

11 Wang, Y. *et al.* Recent progress in nickelate superconductors. *Natl. Sci. Rev.* **12**, nwaf373 (2025).

12 Chen, Z. & Huang, H. The nickelate bridge between cuprate and iron-based superconductivity. *Quantum Front.* **4**, 17 (2025)

13 Li, F. *et al.* Bulk superconductivity up to 96 K in pressurized nickelate single crystals. *Nature* **649**, 871-878 (2026).

14 Li, Q. *et al.* Signature of superconductivity in pressurized $La_4Ni_3O_{10}$. *Chin. Phys. Lett.* **41**, 017401 (2024).

15 Zhang, M. et al. Superconductivity in trilayer nickelate $La_4Ni_3O_{10}$ under pressure. *Phys. Rev. X* **15**, 021005 (2025).

16 Li, F. *et al.* Design and synthesis of three-dimensional hybrid Ruddlesden-Popper nickelate single crystals. *Phys. Rev. Mater.* **8**, 053401 (2024)

17 Luo, Z., Hu, X., Wang, M., Wú, W. & Yao, D.-X. Bilayer two-orbital model of $La_3Ni_2O_7$ under pressure. *Phys. Rev. Lett.* **131**, 126001 (2023).

18 Shen, Y., Qin, M. & Zhang, G.-M. Effective bi-layer model Hamiltonian and density-matrix renormalization group study for the high-$T_c$ superconductivity in $La_3Ni_2O_7$ under high pressure. *Chin. Phys. Lett.* **40**, 127401 (2023).

19 Yang, J. *et al.* Orbital-dependent electron correlation in double-layer nickelate $La_3Ni_2O_7$. *Nat. Commun.* **15**, 4373 (2024).

20 Li, Y. *et al.* Electronic correlation and pseudogap-like behavior of high-temperature superconductor $La_3Ni_2O_7$. *Chin. Phys. Lett.* **41**, 087402 (2024).

21 Au-Yeung, C. C. *et al.* Universal electronic structure of multi-layered nickelates via oxygen-centered planar orbitals. Preprint at https://arxiv.org/abs/2502.20450 (2025).

22 Yang, Y.-f., Zhang, G.-M. & Zhang, F.-C. Interlayer valence bonds and two-component theory for high-$T_c$ superconductivity of $La_3Ni_2O_7$ under pressure. *Phys. Rev. B* **108**, L201108 (2023).

23 Lechermann, F., Gondolf, J., Botzel, S. & Eremin, I. M. Electronic correlations and superconducting instability in $La_3Ni_2O_7$ under high pressure. *Phys. Rev. B* **108**, L201121 (2023).

24 Lu, C., Pan, Z., Yang, F. & Wu, C. Interlayer-coupling-driven high-temperature superconductivity in $La_3Ni_2O_7$ under pressure. *Phys. Rev. Lett.* **132**, 146002 (2024).

25 Fan, Z. *et al.* Superconductivity in nickelate and cuprate superconductors with strong bilayer coupling. *Phys. Rev. B* **110**, 024514 (2024).

26 Jiang, K., Wang, Z. & Zhang, F.-C. High-temperature superconductivity in $La_3Ni_2O_7$. *Chin. Phys. Lett.* **41**, 017402 (2024).

27 Gu, Y., Le, C., Yang, Z., Wu, X. & Hu, J. Effective model and pairing tendency in the bilayer Ni-based superconductor $La_3Ni_2O_7$. *Phys. Rev. B*, **111**, 174506 (2025).

28 Zhao, Y.-F. & Botana, A. S. Electronic structure of Ruddlesden-Popper nickelates: strain to mimic the effects pressure. *Phys. Rev. B* **111**, 115154 (2025).

29 Yue, C. *et al.* Correlated electronic structures and unconventional superconductivity in bilayer nickelate heterostructures. *Natl. Sci. Rev.* 1**2**, nwaf253 (2025).

30 Hu, X., Qiu, W. Chen, C.-Q., Luo, Z. & Yao, D.-X. Electronic structures and multi-orbital models of $La_3Ni_2O_7$ thin films at ambient pressure. *Commun. Phys.* **8**, 506 (2025)

31 H C Regan, B., Bhatta, X. Z., Zhong, Y. & Jia, C. Structural and electronic evolution of bilayer nickelates under biaxial strain. Preprint at https://arxiv.org/abs/2502.01624 (2025).

32 Liu, Y. *et al.* Superconductivity and normal-state transport in compressively strained $La_2PrNi_2O_7$ thin films. *Nat. Mater.* **24**, 1221-1227 (2025).

33 Hao, B. *et al.* Superconductivity and phase diagram in Sr-doped $La_3Ni_2O_7$ thin films. *Nat. Mater.* **24**, 1756-1762 (2025).

34 Li, P. *et al.* Angle-resolved photoemission spectroscopy of superconducting $(La,Pr)_3Ni_2O_7/SrLaAlO_4$ heterostructures. *Natl. Sci. Rev.* **12**, nwaf205 (2025).

35 Shen, J. *et al.* Nodeless superconducting gap and electron-boson coupling in $(La,Pr,Sm)_3Ni_2O_7$ films. Preprint at https://arxiv.org/abs/2502.17831 (2025).

36 Wang, B. Y. *et al.* Electronic structure of compressively strained thin film $La_2PrNi_2O_7$. Preprint at https://arxiv.org/abs/2504.16372 (2025).

37 Sun, W. *et al.* Observation of superconductivity-induced leading-edge gap in Sr-doped $La_3Ni_2O_7$ thin films. Preprint at https://arxiv.org/abs/2507.07409 (2025).

38 Zhou, G. *et al.* Gigantic-oxidative atomic-layer-by-layer epitaxy for artificially designed complex oxides. *Natl. Sci. Rev.* **12**, nwae429 (2024).

39 Chen, X. *et al.* Polymorphism in the Ruddlesden–Popper nickelate $La_3Ni_2O_7$: discovery of a hidden phase with distinctive layer stacking. *J. Am. Chem. Soc.* **146**, 3640-3645 (2024).

40 Dong, Z. *et al.* Interstitial oxygen order and its competition with superconductivity in

$La_2PrNi_2O_{7+\delta}$. *Nat. Mater.* **24**, 1927-1934 (2025).

41 Puphal, P. *et al.* Unconventional crystal structure of the high-pressure superconductor $La_3Ni_2O_7$. *Phys. Rev. Lett.* **133**, 146002 (2024).

42 Huang, C. et al. Superconductivity in monolayer-trilayer phase of $La_3Ni_2O_7$ under high pressure. Preprint at https://arxiv.org/abs/2510.12250v1 (2025).

43 Wang, M. *et al.* Superconducting dome in $La_{3-x}Sr_xNi_2O_{7-\delta}$ thin films. *Phys. Rev. Lett.* **136**, 066002 (2026).

44 Wang, H. *et al.* Electronic structures across the superconductor-insulator transition at $La_{2.85}Pr_{0.15}Ni_2O_7/SrLaAlO_4$ interfaces. Preprint at https://arxiv.org/abs/2502.18068 (2025).

45 Zhang, Y. *et al.* Electronic structure, and magnetic and superconducting pairing tendencies of the alternating single layer-bilayer stacking nickelate $La_5Ni_3O_{11}$ under pressure. *Phys. Rev. B* **112**, 094515 (2025).

46 LaBollita, H. & Botana, A. S. Correlated electronic structure of the alternating monolayer-bilayer nickelate $La_5Ni_3O_{11}$. *Phys. Rev. B* https://doi.org/10.1103/k83c-qgx1 (2026).

47 Zhang, M., Chen, C.-Q., Yao, D.-X. & Yang, F. Pairing mechanism and superconductivity in pressurized $La_5Ni_3O_{11}$. *Sci. China Phys. Mech. Astron.* https://www.sciengine.com/doi/10.1007/s11433-025-2907-0 (2026).

48 Ouyang, Z., He, R.-Q. & Lu, Z.-Y. Phase diagrams and two key factors to superconductivity of Ruddlesden-Popper nickelates. *Phys. Rev. B* **112**, 045127 (2025).

49 Abadi, S. N. *et al.* Electronic structure of the alternating monolayer-trilayer phase of $La_3Ni_2O_7$. *Phys. Rev. Lett.* **134**, 126001 (2025).

50 Cao, Y.-H., Jiang, K.-Y., Lu, H.-Y., Wang, D. & Wang, Q.-H. Strain-engineered electronic structure and superconductivity in $La_3Ni_2O_7$ thin films. *Sci. China Phys. Mech. Astron.* **69**, 247412 (2026).

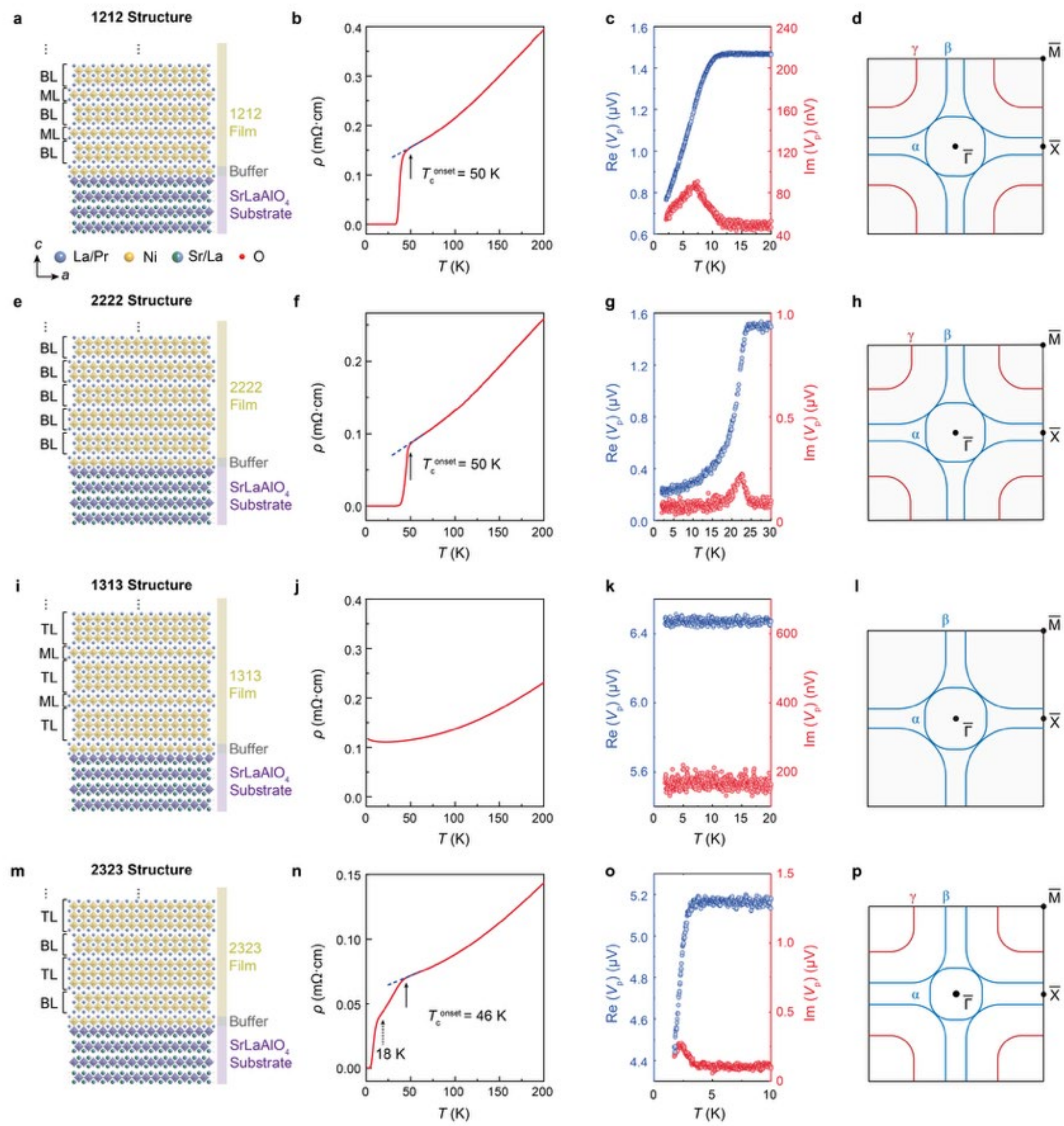

**Figure 1| Comparison of 1212, 2222, 1313, and 2323 nickelate thin films. a**, **e**, **i**, **m**, Schematics of the 1212 (a), 2222 (e), 1313 (i), and 2323 (m) films grown on $SrLaAlO_4$ substrates. ML, BL, and TL denote monolayer, bilayer, and trilayer blocks, respectively. **b**, **f**, **j**, **n**, Temperature-dependent resistivity ($\rho$-$T$) curves for the corresponding films. $T_{c,onset}$ is defined as the deviation point from the normal-state linear fit. Note that the resistivity of our 2222 sample (f) is comparable to the films reported previously[32]. Film thicknesses are 5 nm (1212), 7 nm (2222), 8 nm (1313), and 8 nm (2323). **c**, **g**, **k**, **o**, Diamagnetic responses of the same sample in **b**, **f**, **j**, **n**, measured via two-coil mutual inductance. Blue and red dots represent the real (Re($V_p$)) and imaginary (Im($V_p$)) components of the pickup coil voltage, respectively. These samples are of pure phases without capping layers. **d**, **h**, **l**, **p**, Schematics of the Fermi surfaces determined by ARPES. Blue and red lines indicate α/β and γ pockets, respectively.

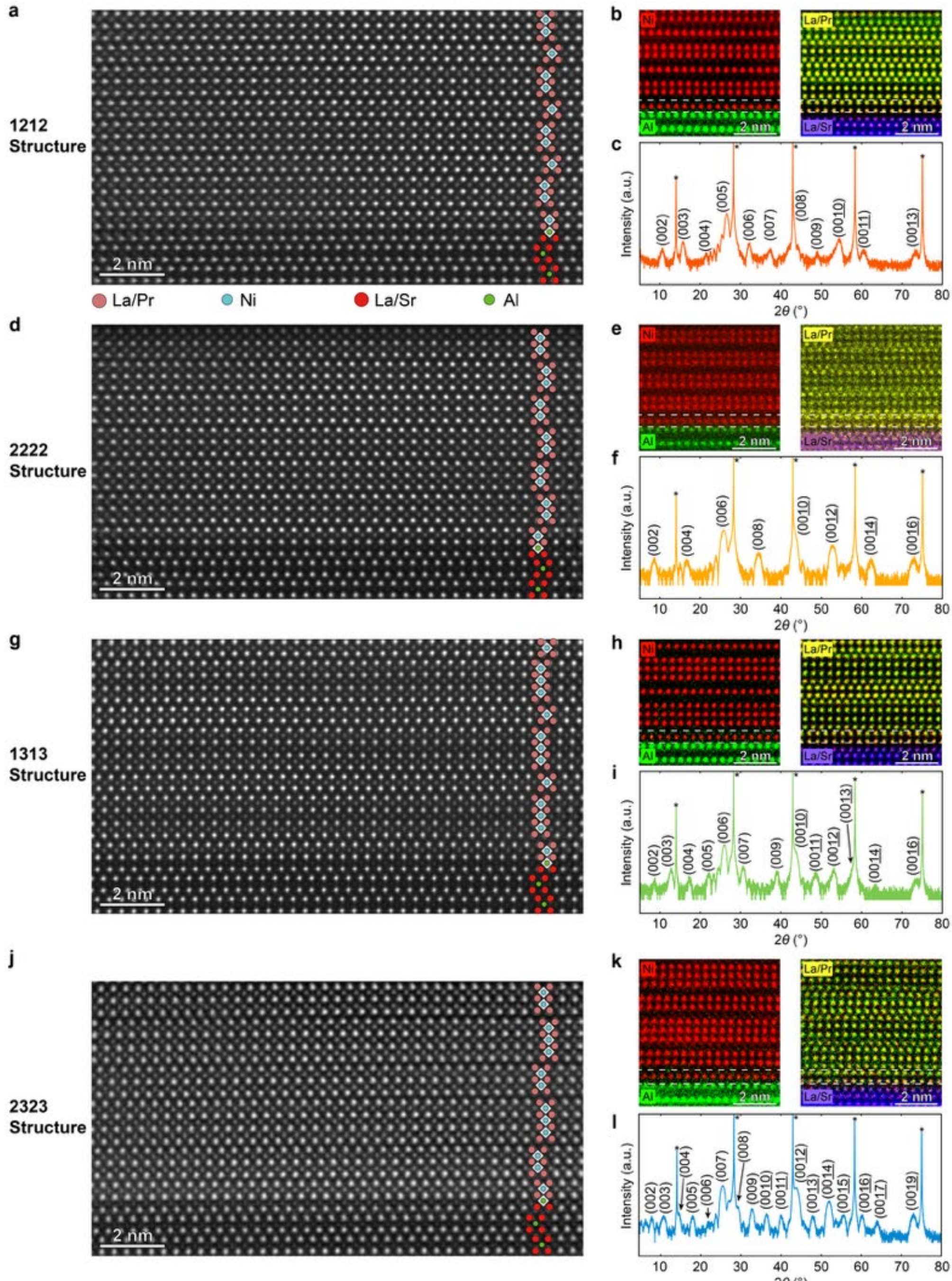

**Figure 2 | Structural characterizations of 1212, 2222, 1313, and 2323 nickelate thin films. a**, **d**, **g**, **j**, Scanning transmission electron microscopy (STEM) high-angle annular dark-field (HAADF) images of the 1212 (a), 2222 (d), 1313 (g), and 2323 (j) films grown on $SrLaAlO_4$ substrates. **b**, **e**, **h**, **k**, Atomically resolved energy dispersive spectroscopy (EDS) elemental maps (Ni, Al, La, Pr, Sr) corresponding to the films shown in the left panels. White dashed lines indicating film, buffer layer, and substrate boundaries. **c**, **f**, **i**, **l**, Out-of-plane X-ray diffraction (XRD) patterns for the 1212 (c), 2222 (f), 1313 (i), and 2323 (l) films. Film diffraction peaks are indexed as (00*l*), and substrate peaks are marked with asterisks. Note that samples analyzed by STEM were capped with ~2-unit-cell $SrTiO_3$ for protection, whereas samples for XRD measurements were uncapped.

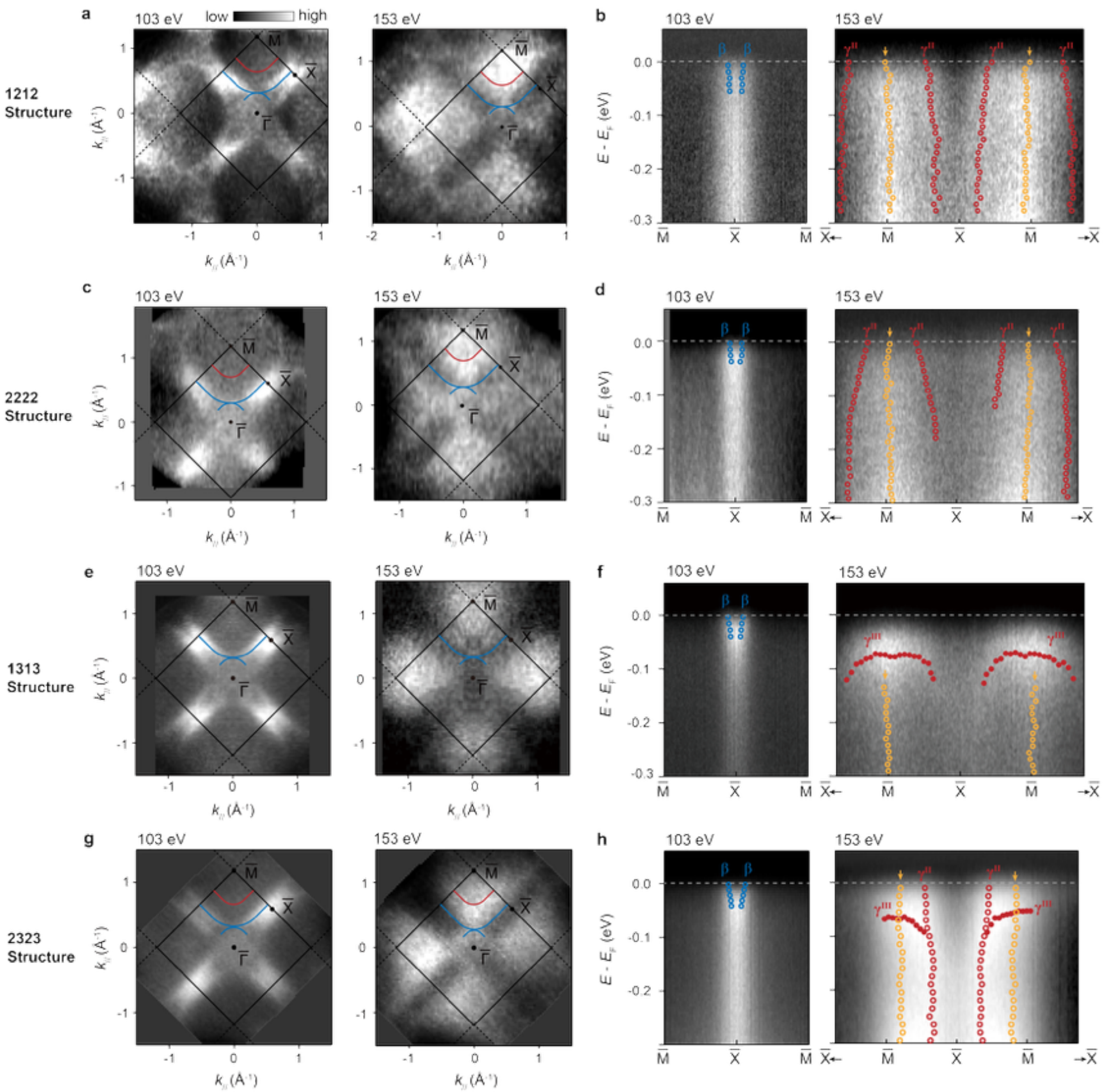

**Figure 3 | Electronic band structures of 1212, 2222, 1313, and 2323 nickelate thin films.** **a**, **c**, **e**, **g**, Fermi surface maps measured using 103 eV and 153 eV photons for the 1212 (**a**), 2222 (**c**), 1313 (**e**), and 2323 (**g**) films. The integration window is 50 meV around $E_F$, with schematic band contours overlaid. The surface terminations are bilayer for the 1212 and 2323 films, and trilayer for the 1313 film. **b**, **d**, **f**, **h**, ARPES spectral cuts along $\bar{M}$-$\bar{X}$-$\bar{M}$ corresponding to the films in the left panels. Blue circles denote peak positions of β bands. Red symbols denote peak positions of γ bands (including $\gamma^{II}$ and $\gamma^{III}$ in h). Yellow circles mark the vertical feature near zone corner. Peak positions are extracted from momentum distribution curve (MDC) fits (open circles). For 1313 and 2323 films (f and h), band dispersions derived from energy distribution curve (EDC) fits are shown with red dots. The color bar indicates spectral intensity. Film thicknesses are 7 nm (1212), 5 nm (2222), 8 nm (1313), and 6 nm (2323). These samples are of pure phases without capping layers.

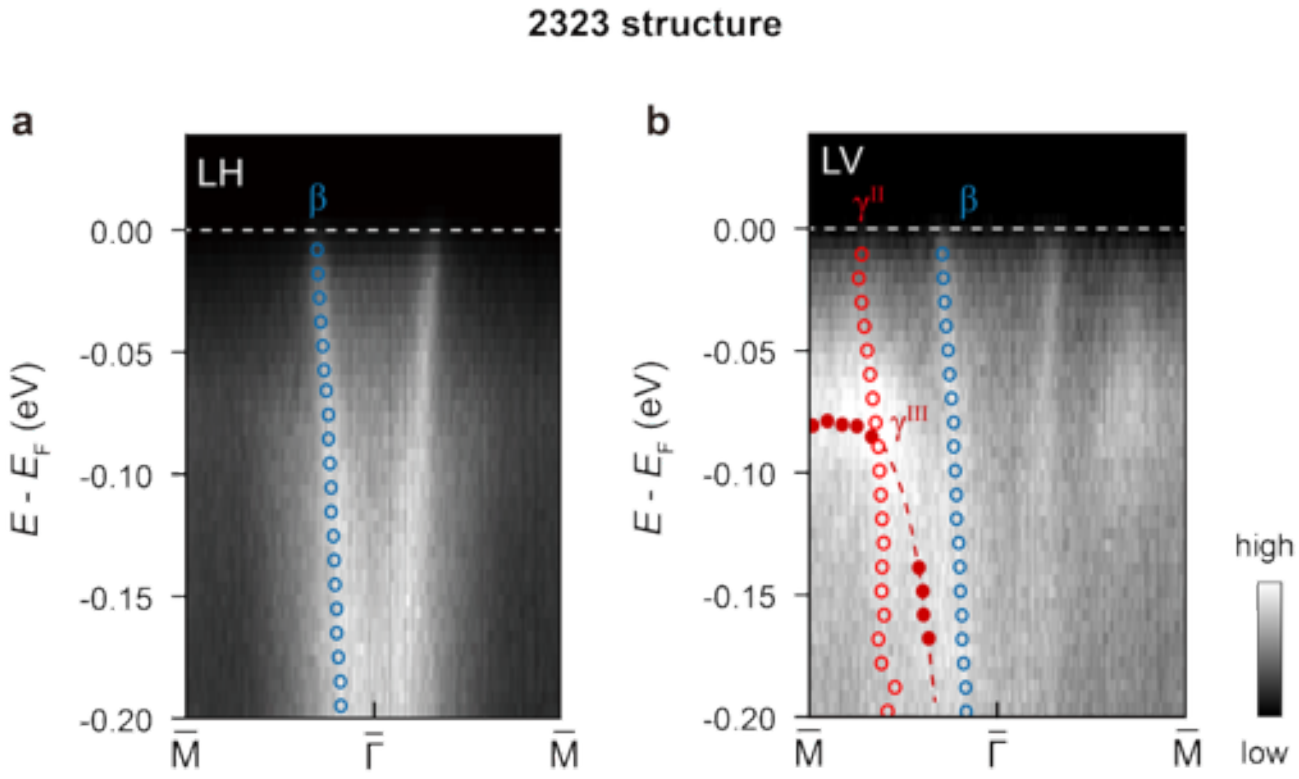

**Figure 4 | Polarization dependence of 2323 structure thin film at 103 eV photons.** The ARPES spectra cuts along $\bar{M}$-$\bar{\Gamma}$-$\bar{M}$ measured by linear horizontal (LH) (**a**) and linear vertical (LV) (**b**) 103 eV photons. The film is 5 nm thick with trilayer termination. The blue circles, red circles and red dots represent the peak positions from α/β, $\gamma^{II}$ and $\gamma^{III}$ bands, respectively. The color bar indicates spectral intensity.

| | **1212** | **2222** | **1313** | **2323** |
|---|---|---|---|---|
| | superconducting | superconducting | non-superconducting | superconducting |
| **$\gamma^{II}$ forming underlying Fermi pocket** | √ | √ | | √ |
| **$\gamma^{III}$ lying below Fermi level** | | | √ | √ |

Table 1 Key ARPES features and superconducting properties of RP superstructures

## Methods

**Thin film growth.** All samples were prepared on as-received $SrLaAlO_4$ (001) substrates (MTI-Kejing), using the gigantic-oxidative atomic layer-by-layer epitaxy (GAE) method[38]. During growth, $La_{0.67}Pr_{0.33}O_x$ and $NiO_y$ targets were alternately ablated. The ablation sequences for the 1212 and 1313 superstructures followed the stacking order of the target crystal structures. For example, the growth sequence of the 1212 superstructure is [(La,Pr)O-$NiO_2$-(La,Pr)O]-[(La,Pr)O-$NiO_2$-(La,Pr)O-$NiO_2$-(La,Pr)O]. Prior to the growth of the 2222 and 2323 structures, a buffer layer with the sequence (La,Pr)O-$NiO_2$-(La,Pr)O was first deposited to mitigate interfacial defects and stabilize the layered architecture during subsequent deposition[51,52]. The stoichiometry was controlled by precisely adjusting the number of laser pulses required to deposit one atomic layer. Calibration of the pulse numbers for the two targets was performed by synthesizing 3-unit-cell (UC) $(La,Pr)_3Ni_2O_7$ films on $SrLaAlO_4$ substrates. The XRD patterns of the calibration sample include at least the (004) diffraction peak, indicating the formation of a long-range bilayer phase, which ensures near-stoichiometric growth when synthesizing hybrid Ruddlesden-Popper structures. Typically, 100-150 pulses were used for ablating $La_{0.67}Pr_{0.33}O_x$ and $NiO_y$ targets, achieving stoichiometric precision better than 1%. The precise control of the cation stoichiometry is important to grow the pure RP bilayer phase, which is prone to the intergrowth of secondary phases such as trilayer phase[5,41,53]. See another report for the details of the sample optimization[52].

The $La_{0.67}Pr_{0.33}O_x$ and $NiO_y$ targets were synthesized from a stoichiometric mixture of $Pr_6O_{11}$ and $La_2O_3$ powders and NiO powder, respectively. $La_{0.67}Pr_{0.33}O_x$ target was sintered twice, each time for 6 h at 1100 °C and $NiO_y$ target was sintered once for 6 h at 1100 °C. Film growth was carried out at temperatures ranging from 760 to 850 °C under a mixed atmosphere of purified ozone and oxygen. The total growth pressure was maintained at 10 Pa, with the ozone partial pressures ranging from 1 Pa to 3 Pa. The substrate temperature is measured by an infrared pyrometer from the back side of the Inconel600 sample holder, typically 70-80 °C higher than the temperature measured from the sample surface. The laser fluence was set to 1.4-1.7 $J \cdot cm^{-2}$, and the pulsed laser repetition rate was 4 Hz. After deposition, the samples were cooled at a rate of 100 °C $min^{-1}$ to below 100 °C before being transferred from the growth chamber to the load-lock chamber, in order to prevent oxygen loss.

**Transport and mutual inductance measurements.** Pt Hall-bar electrodes were deposited onto the samples via magnetron sputtering and bonded to sample holders using aluminum wires with an ultrasonic wire bonder. Transport measurements were performed in pulse tube cryostats. A superconducting magnet was used to apply the magnetic field. The in-plane magnetic field orientation was calibrated by identifying the rotation angle at which the resistance reached its minimum under a fixed magnetic field and temperature. Mutual inductance measurements were carried out using the same setup as described previously[6]. In Extended Data Figure 5, the temperature-dependent critical fields were analyzed using the Ginzburg-Landau model[54], except for the $B_c^{50\%}$ data of the 2323 structure, which required the use of Gurevich's two-band model to capture its two-step feature[55]. The coherence length was estimated via the equation: $\xi_0 = \sqrt{(\Phi_0/2\pi B_{c2})}$, where $\Phi_0$ denotes the magnetic flux quantum. The angular dependence of $T_c$ was well described using the Tinkham model[56], with a two-component piecewise fitting applied to the 2323 structure.

**XRD and STEM measurements.** X-ray diffraction (XRD) $\theta$-$2\theta$ symmetric scans were performed using an X-ray diffractometer (SmartLab, Rigaku Corporation). Cross-sectional STEM specimens were prepared with a FEI Helios G4 HX dual-beam focused ion beam (FIB) and scanning electron microscope (SEM) machine. HAADF-STEM imaging was carried out on a FEI Themis Z operated at 200 kV, equipped with a Cs Probe Corrector (DCOR), a high-brightness field-emission gun (X-FEG) and a monochromator. Imaging conditions included inner and outer collection angles ($\beta_1$ and $\beta_2$) of 90 and 200 mrad, a semiconvergence angle of 25 mrad, and beam currents of ~40 pA for HAADF imaging and 136 pA for EDS chemical analysis. EDS measurements of $(La,Pr)_3Ni_2O_7$ were performed using the Super-X FEI System in STEM mode. EDS data for $(La,Pr)_5Ni_3O_{11}$, $(La,Pr)_6Ni_4O_{14}$ and $(La,Pr)_7Ni_5O_{17}$ were acquired on a JEOL ARM200F, using the Super Dual EDS System.

**ARPES measurements and analysis.** The ARPES measurements were performed at beamline BL03U of Shanghai Synchrotron Radiation Facility (SSRF) in China. The energy resolution was better than 10 meV at photon energy of 100 eV. Base pressure of the beamline station was lower than $7\times10^{-11}$ Torr. Immediately after growth, the samples are transferred to a cryogenic ultrahigh-vacuum suitcase (base pressure lower than $5\times10^{-10}$ Torr, base temperature ~100 K), such that the samples are rapidly cooled from growth temperature to below 200 K. During the whole transfer process from growth chamber to the ARPES endstation, the samples were maintained below 200 K to avoid surface oxygen loss and to ensure the measurement of intrinsic superconducting state. During the measurement, the sample holder temperature is kept below 15 K. Another challenge for ARPES studies is the thermodynamic instability of the RP bilayer phase, which can form trilayer surface impurities. These impurities are undetectable by transport but pose risks of misinterpretation for ARPES, as their higher photoemission intensity allows them to be selected during real-space sample scanning. The phase purity and oxygen homogeneity are further evidenced by the uniform spectral intensity observed during our real-space scanning of our films. The methods for ARPES spectra fitting and background subtraction are presented in Extended Data Figures 8. It is noted that the waterfall-like features, nearly vertical and smearing out in dispersion, are commonly observed in our results, which is also reported in both infinite-layer nickelate and cuprates[57-60]. While different photon energies are employed in this study, the significant intensity variation cannot be rationalized purely in terms of $k_z$ dispersion. The effect is instead attributed to the energy-dependent photoemission cross-sections of different orbitals[61].

51 Kim, J. *et al.* Defect engineering in $A_2BO_4$ thin films via surface-reconstructed $LaSrAlO_4$ substrates. *Small Method* **6**, 2200880 (2022).

52 Lyu, W. *et al.* Preparation and optimization of high-temperature superconducting Ruddlesden-Popper nickelate thin films. *Acta Phys. Sin.* **74**, 227403 (2025).

53 Shi, Y. *et al.* Critical thickness and long-term ambient stability in superconducting $LaPr_2Ni_2O_7$ films. *Adv. Mater.* **38**, e10394 (2026).

54 Harper, F. E. & Tinkham, M. The mixed state in superconducting thin films. *Phys. Rev.* **172**, 441-450 (1968).

55 Gurevich, A. Enhancement of the upper critical field by nonmagnetic impurities in dirty two-gap superconductors. *Phys. Rev. B* **67**, 184515 (2003).

56 Tinkham, M. *Introduction to superconductivity*. (McGraw-Hill, Inc., 1996).

57 Ding, X. *et al.* Cuprate-like electronic structures in infinite-layer nickelates with substantial hole dopings. *Natl. Sci. Rev.* **11**, nwae194 (2024).

58 Sun, W. *et al.* Electronic structure of superconducting infinite-layer lanthanum nickelates. *Sci. Adv.* **11**, eadr5116 (2025).

59 Graf, J., Gweon, G. H. & Lanzara, A. Universal waterfall-like feature in the spectral function of high temperature superconductors. *PHYSICA C* **460-462**, 194-197 (2007).

60 Krsnik, J. & Held, K. Local correlations necessitate waterfalls as a connection between quasiparticle band and developing Hubbard bands. *Nat. Commun.* **16**, 255 (2025).

61 Bansil, A., Markiewicz, R. S., Kusko, C., Lindroos, M. & Sahrakorpi, S. Matrix element and strong electron correlation effects in ARPES from cuprates. *J. Phys. Chem. Solids* **65**, 1417-1421 (2004).

## Acknowledgements

We acknowledge discussions with J. He, M. Wang, H. Yuan, D.-X. Yao and G.-M. Zhang. This work is supported by the National Key Research and Development

Program of China (Grant Nos. 2024YFA1408100, 2022YFA1403101, 2023YFA1406304), the National Natural Science Foundation of China (Grant Nos. 92565303, 92265112, 12374455, 52388201, 12504079, 12504165, 12504166, 12504161, 125B2072, 12494593, 52473302, 12461160252, T2525009), Guangdong Major Project of Basic Research (2025B0303000004), the Quantum Science Strategic Initiative of Guangdong Province (Grant Nos. GDZX2501001, GDZX2401004, GDZX2201001, GDZX2301006), the Municipal Funding Co-Construction Program of Shenzhen (Grant Nos. SZZX2401001, SZZX2301004), and the Science and Technology Program of Shenzhen (Grant No. KQTD20240729102026004). Dawei Shen acknowledges the support from Anhui Provincial Natural Science Foundation (No. 2408085J003). Junhao Lin acknowledges the support from Natural Science Foundation of Guangdong Province, China (Grant Nos. 2023B1515120039). Yueying Li acknowledges the support by China Postdoctoral Science Foundation (Grants Nos. GZC20240649, 2024M761276). Zhicheng Jiang acknowledges the China National Postdoctoral Program for Innovative Talents (BX20240348) and Xiaomi Young Talents Program. We thank the Shanghai Synchrotron Radiation Facility (SSRF) of BL03U (31124.02.SSRF.BL03U) for the assistance on ARPES measurements. We acknowledge the support from the International Station of Quantum Materials.

**Author contributions**

Q.-K.X. and Z.C. supervised the entire project. Z.C. initiated the study and coordinated all the research efforts. Z.N. and W.L. performed thin-film growth, with assistance from Y.C., under supervision from G.Z. and Z.C. Y.L., L.X and Z.J. conducted the ARPES measurements, with assistance from W.S., under supervision from Z.C., P.L., and D.S. D.S. provided access to synchrotron ARPES facility. P.F., Z.N., and H.H. conducted STEM and EDS measurements and analysis, under supervision of J.L. and Z.C. Z.N. and H.W. performed the transport and mutual inductance measurements. J.-F.J. and all other authors participated in discussions. Z.C., Y.L., Z.N., P.L. wrote the manuscript with input from all other authors. Z.N., Y.L., W.L., L.X., and Z.J. contributed equally.

**Data availability**

Source data are provided with this paper.

**Competing interests** The authors declare no competing interests.

**Correspondence and requests for materials** should be addressed to Zhuoyu Chen, Qi-Kun Xue, Peng Li, or Dawei Shen.

**Reprints and permissions information** is available at http://www.nature.com/reprints.

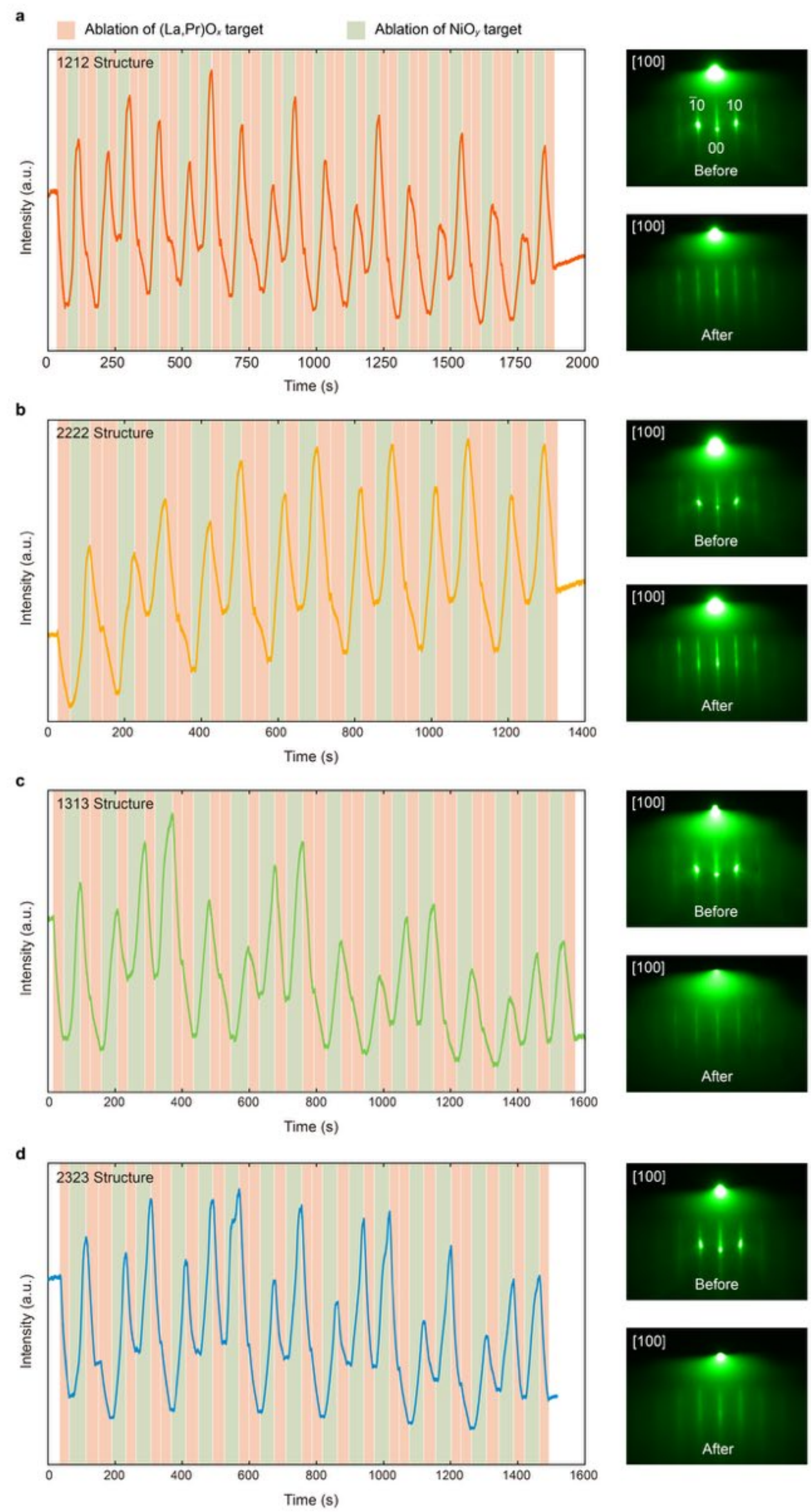

**Extended Data Figure 1 | Reflective high-energy electron diffraction (RHEED).** RHEED oscillations and the corresponding RHEED patterns before and after growth of the 1212 (**a**), 2222 (**b**), 1313 (**c**), and 2323 (**d**) films. The growth sequence follows the exact stacking sequence of the target structure. For example, the growth sequence of 1212 is: [(La,Pr)O-NiO$_2$-(La,Pr)O]-[(La,Pr)O-NiO$_2$-(La,Pr)O-NiO$_2$-(La,Pr)O].

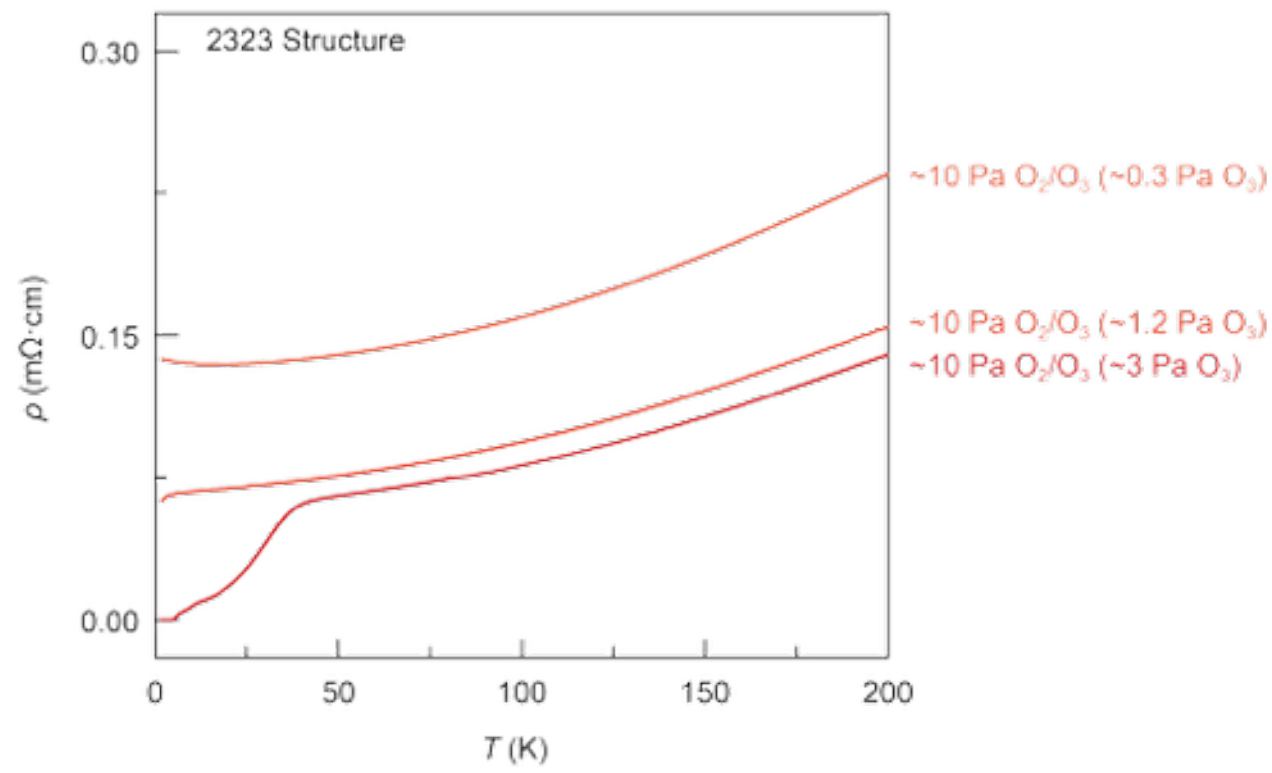

**Extended Data Figure 2 | Impact of oxidant conditions during growth on the electrical transport property of 2323 films.** The growth temperature is controlled to be 800℃ in these experiments.

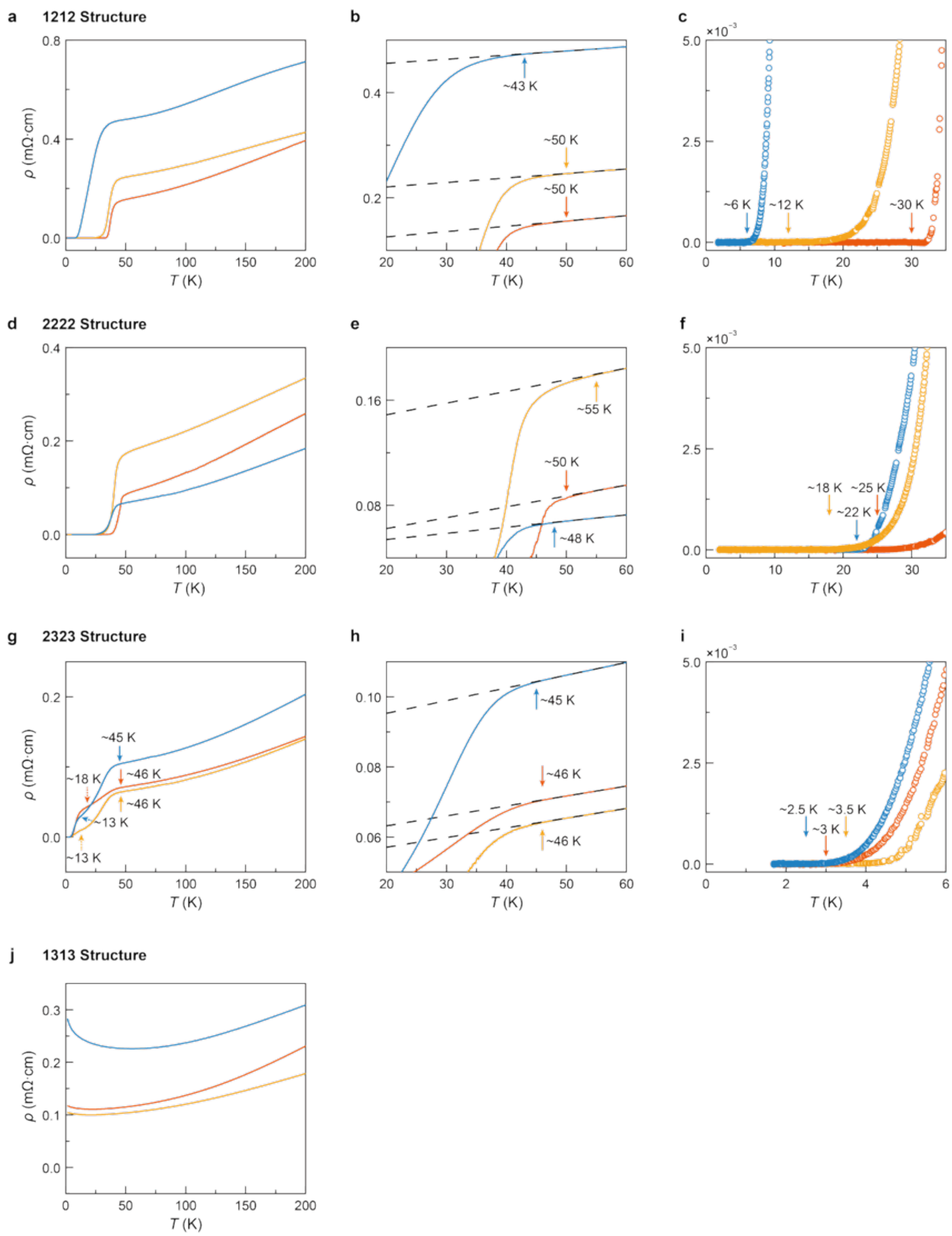

**Extended Data Figure 3 | Temperature-dependent resistivities**. **a-c**, $\rho$-$T$ curves of three 1212 films, and enlarged views around onset and near-zero regions, respectively. Dashed black lines are linear fits to the normal-state resistivity. Arrows indicate the $T_{c,onset}$ and the temperatures at which resistivity drops below noise level. **d-f**, $\rho$-$T$ curves of three 2222 films. **g-i**, $\rho$-$T$ curves of three 2323 films. The second transition is denoted by dashed arrows in **g**. The temperature of the second transition is defined as the temperature at which the temperature derivative of the resistivity ($d\rho/dT$) exhibits a second rise upon cooling. **j**, $\rho$-$T$ curves of three 1313 films.

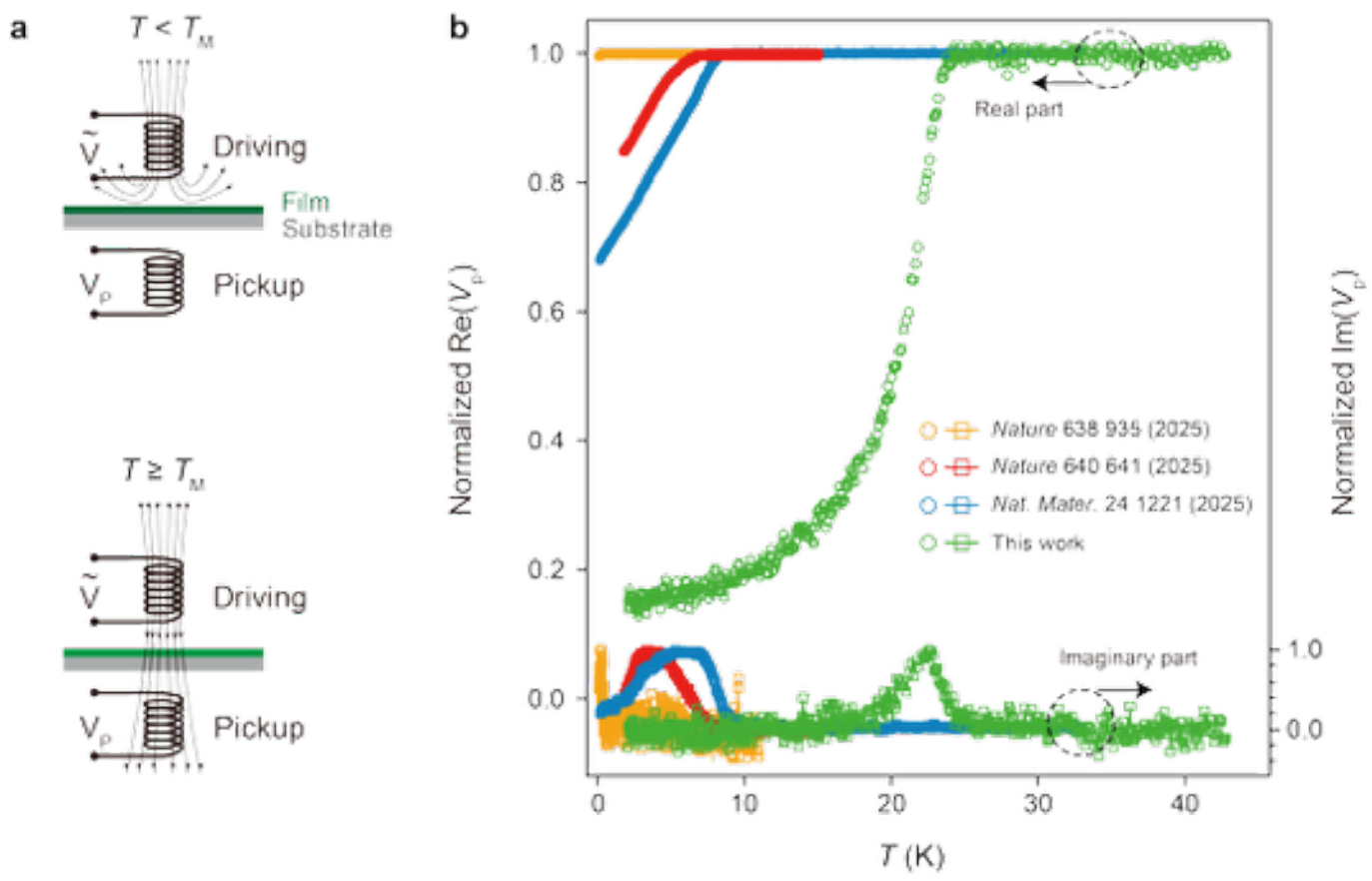

**Extended Data Figure 4 | Comparison of diamagnetic response of the 2222 structure in this work with previous reports. a**, schematic of the two-coil mutual inductance setup. **b**, Normalized real part (Re($V_p$)) and imaginary part (Im($V_p$)) of pickup coil voltage as functions of temperature. The data are from the $La_3Ni_2O_7$ film in ref. 5, $La_{2.85}Pr_{0.15}Ni_2O_7$ film in ref. 6, $La_2PrNi_2O_7$ film in ref. 32 and $La_2PrNi_2O_7$ film in this work, respectively.

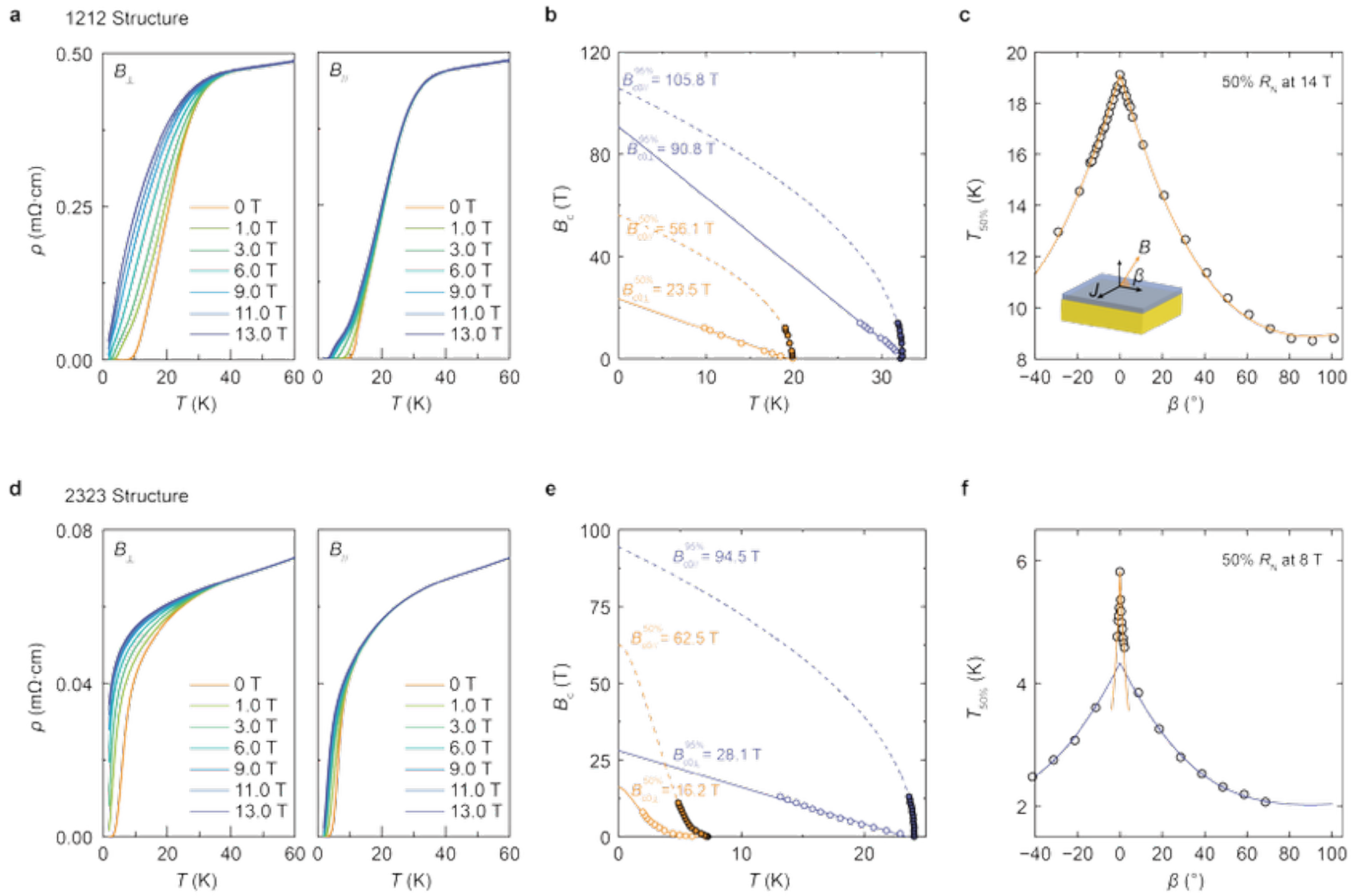

**Extended Data Figure 5 | Magnetic response of 1212 and 2323 nickelate thin films.** **a,** $\rho$-$T$ curves of a superconducting 1212 film measured with magnetic fields applied perpendicular (left panel) and parallel (right panel) to the film plane. **b,** Temperature-dependent in-plane (solid circles) and out-of-plane (open circles) critical magnetic fields. Critical fields are extracted using both $T_{c,95\%}$ (purple) and $T_{c,50\%}$ (orange) corresponding to the temperatures at which the resistance drops to 95% and 50% of the normal-state resistance $R_N$, respectively. The lines represent corresponding fittings of in-plane (dashed lines) and out-of-plane (solid lines) critical magnetic fields using Ginzburg-Landau (GL) model. The fittings yield the in-plane coherence length $\xi_0^{\parallel}$ to be 1.76 nm and 2.42 nm for 95% and 50% critical field, respectively, with the corresponding out-of-plane coherence length $\xi_0^{\perp}$ being 1.90 nm and 3.74 nm. **c,** Angular dependence of the critical temperature $T_{c,50\%}$, defined as the temperature at which the resistance drops to 50% $R_N$ under an applied magnetic field of 14 T. The inset illustrates the measurement geometry, where $\beta$ denotes the angle between the applied field and the film plane. Open circles are the experimental data. The orange curve represents the fitting using the Tinkham model. **d,** Corresponding $\rho$-$T$ curves for a superconducting 2323 film under perpendicular and parallel fields. **e**, Corresponding temperature-dependent in-plane (solid circles) and out-of-plane (open circles) critical magnetic fields for a superconducting 2323 film. The $B_{c,50\%}$ data exhibits an obvious two-step feature and Gurevich's two-band model is used for fitting. The fittings yield the in-plane coherence length $\xi_0^{\parallel}$ to be 1.87 nm and 2.29 nm for 95% and 50% critical field, respectively, with the corresponding out-of-plane coherence length $\xi_0^{\perp}$ being 3.42 nm and 4.51 nm. **h**, Corresponding $T_{c,50\%}$-$\beta$ curve for 2323 structure under an applied magnetic field of 8 T. The dataset with clear two-step feature is separated into two angular regimes and captured by piecewise fits.

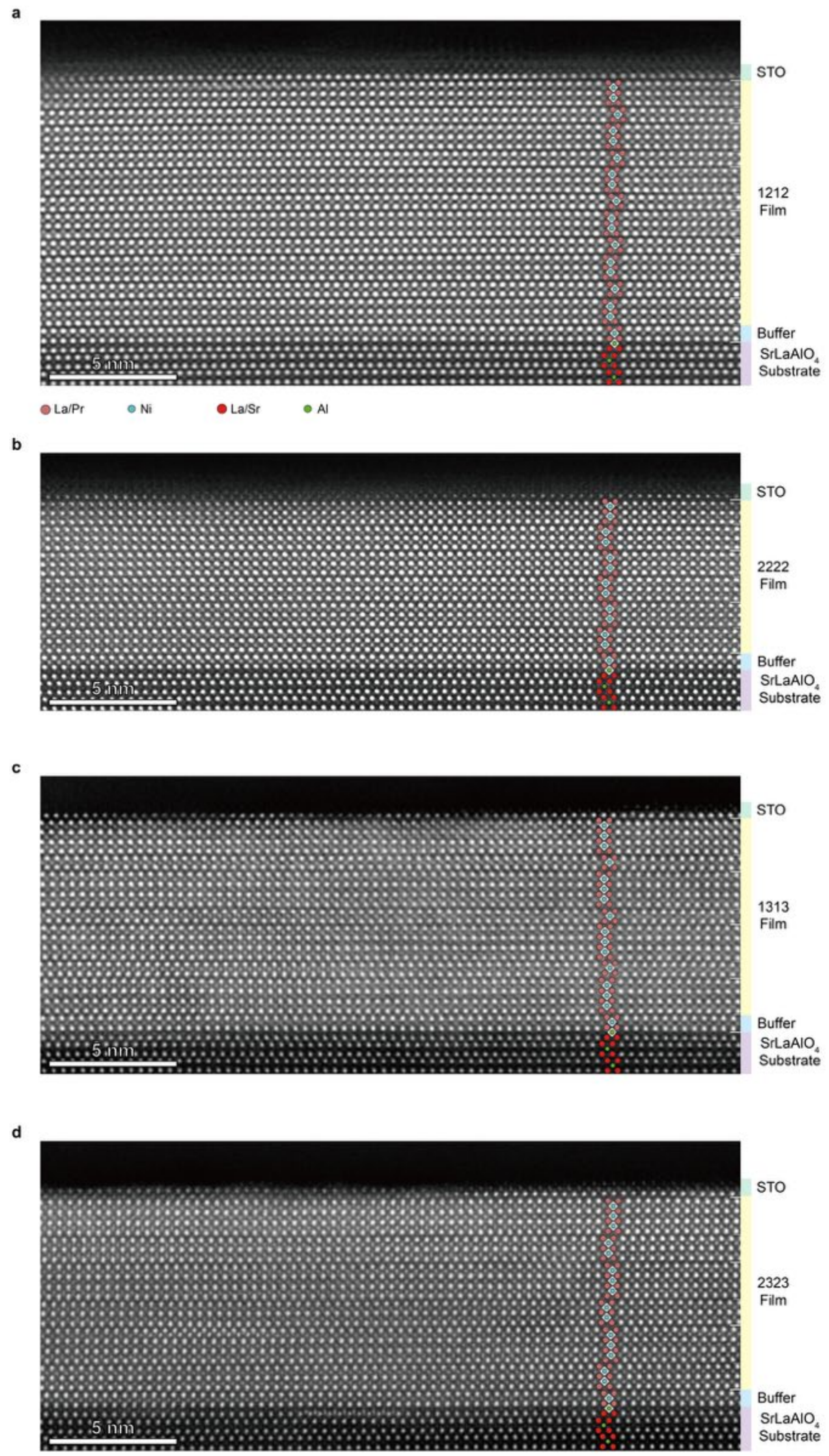

**Extended Data Figure 6 | HAADF images of the 1212 (a), 2222 (b), 1313 (c) and 2323 (d) films on $SrLaAlO_4$ substrates with a large field of view**. All samples prepared for STEM were capped with STO to prevent damage during focused ion beam (FIB) processing. Scale bars, 5 nm.

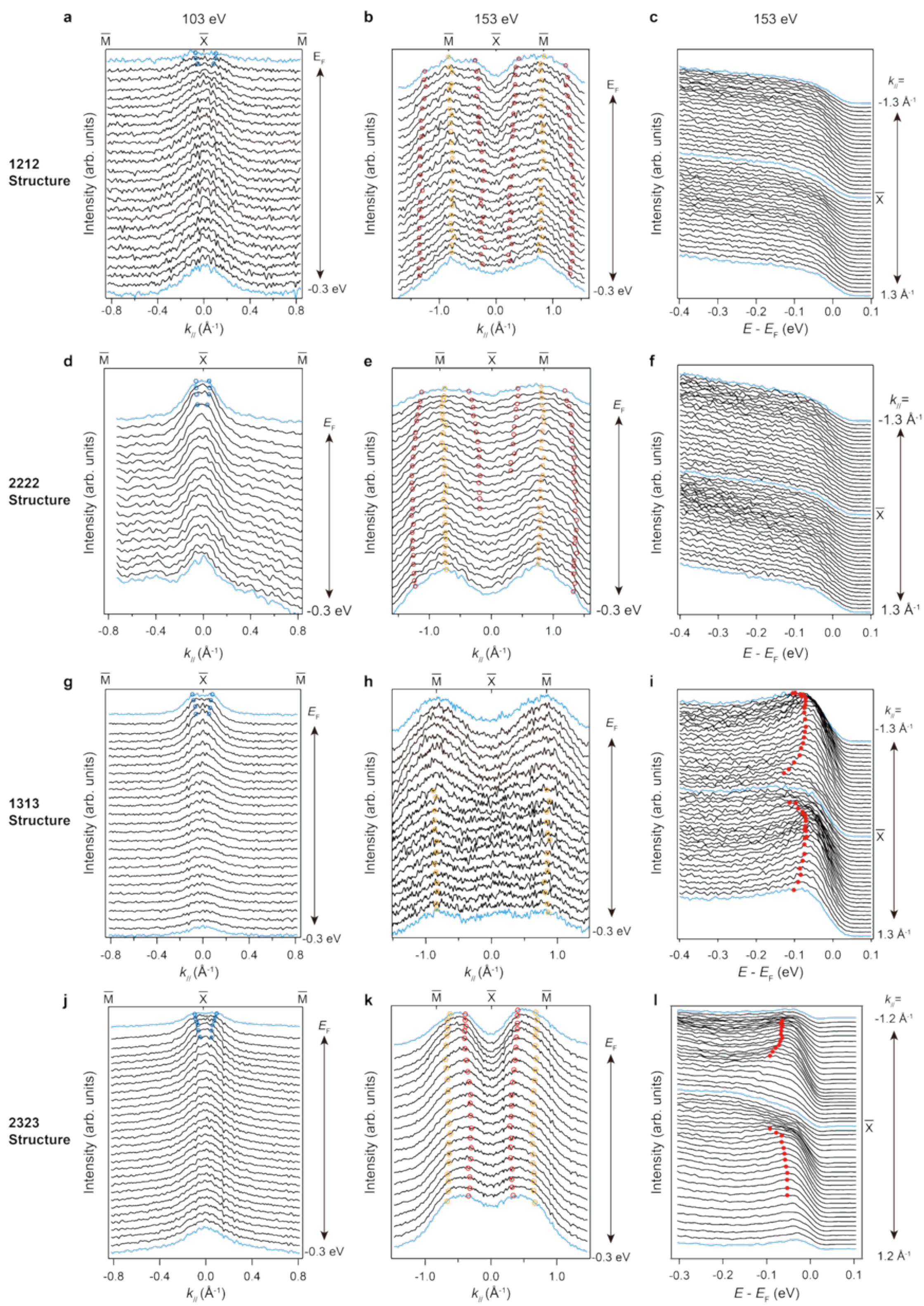

**Extended Data Figure 7 | Momentum and energy distribution curves (MDCs and EDCs) of the ARPES spectral cuts in Figure 3. a**, MDCs corresponding to the spectral cut measured by 103 eV photons in Fig. 3b for 1212 structure. The blue circles mark the peak positions from β bands, extracted from the Lorentzian-peak fits. **b, c**, MDCs

and EDCs corresponding to the spectral cut measured by 153 eV photons in Fig. 3b for 1212 structure. The red and yellow circles represent the peak positions from γ band and $\overline{\mathrm{M}}$-vertical feature, respectively, extracted from the six-Lorentzian-peak fits shown in Extended Data Fig. 8a. The EDCs display a typical Fermi-Dirac shape without obvious peaks near the Fermi level. **d-f**, Corresponding MDCs and EDCs for 2222 structure, which are similar as 1212 structure. **g-i**, Corresponding MDCs and EDCs for 1313 structure. The 153-eV spectrum shows significant differences with 1212 and 2222 structures. The MDCs exhibit the two-lobed lineshape near the Fermi level due to the flat γ band. And the EDCs of 153 eV spectrum show evident peaks under the Fermi level on the background, which can be captured by the Lorentzian fitting with Fermi-Dirac-type background shown in Extended Data Fig. 8b. Note that the intensity scales of MDCs and EDCs have been adjusted to highlight the peaks positions.

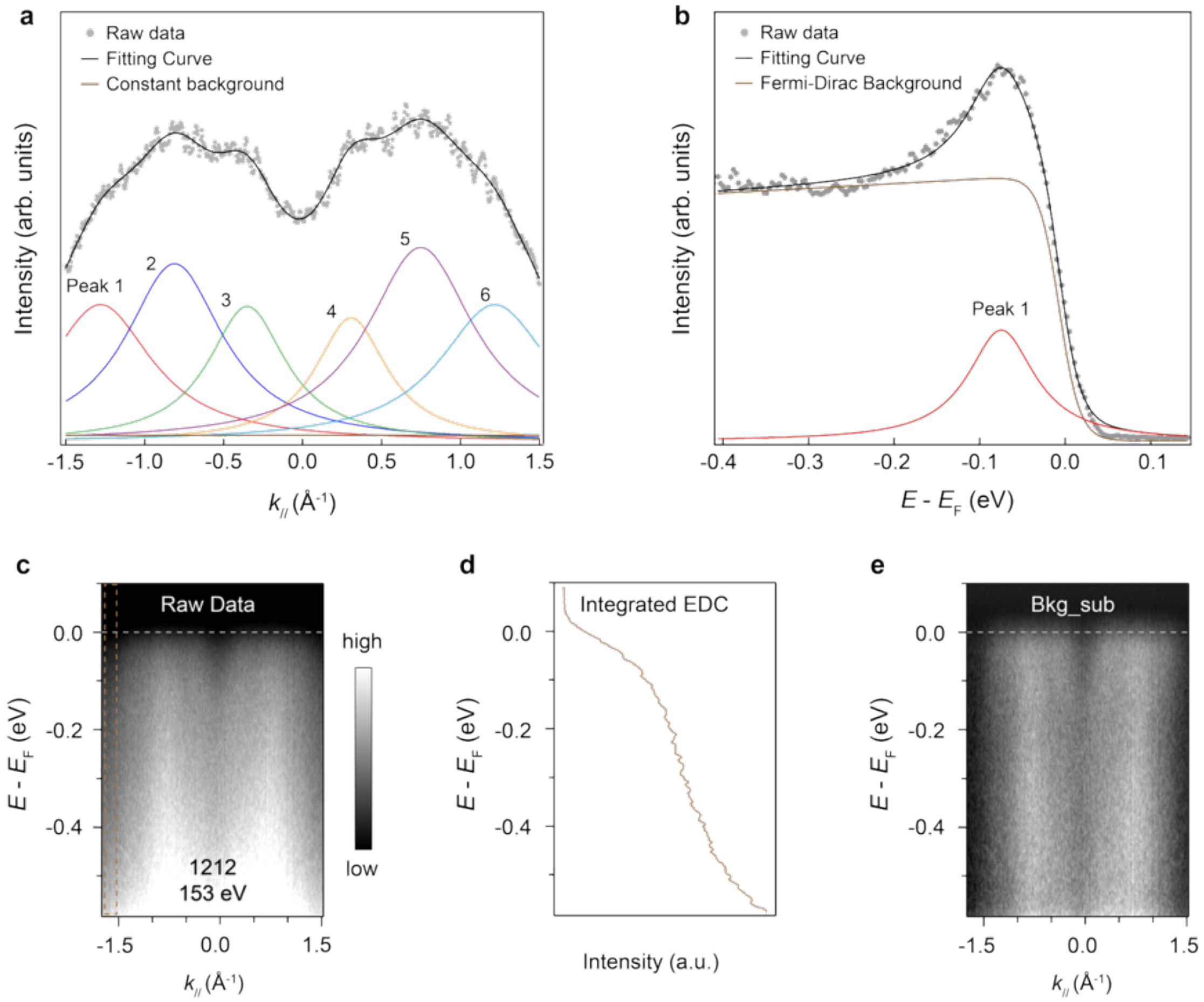

**Extended Data Figure 8 | Fitting methods and background substraction of the ARPES spectra. a,** the multi-Lorentzian-peak fit with the constant background to the six-lobe shape MDC along $\bar{M}$-$\bar{X}$-$\bar{M}$ cut for 1212 hybrid superstructure. **b,** the Lorentzian peak fit with the Fermi-Dirac-type background to the EDC for 1313 superstructure. **c,** The raw data of ARPES spectra along $\bar{M}$-$\bar{X}$-$\bar{M}$ direction for 1212 hybrid structure measured by 153 eV photons. **d,** The corresponding integrated EDC far away from the features as the background. The integrated range is indicated by the brown dashed square in **c**. **e,** The spectra after subtracting the EDC background in **d** to enhance features, same data as the Fig. 3b. This process has been done in Figs. 3b, 3d and Extended Data Fig. 10c for clarity. Other spectra are shown as raw data.

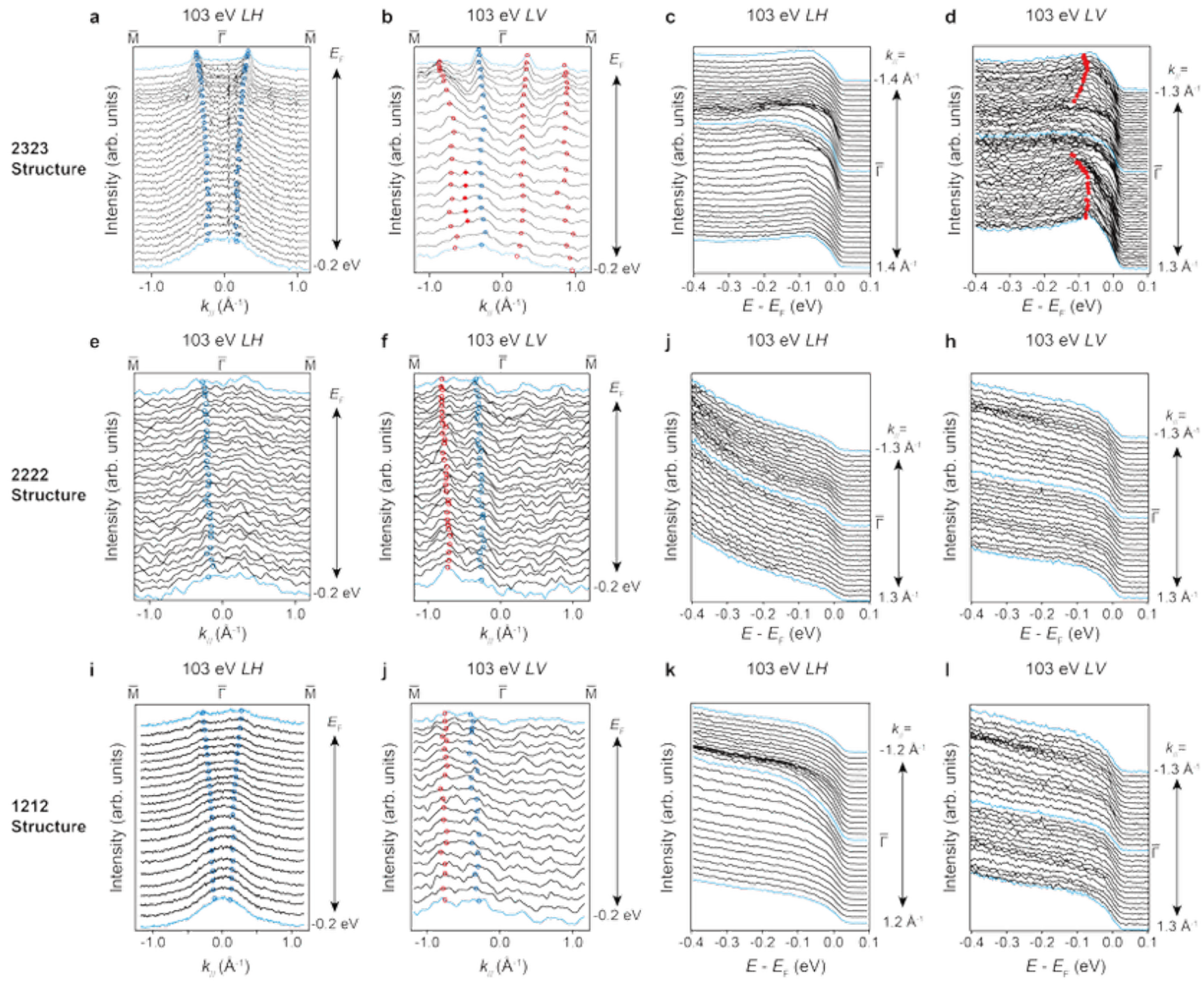

**Extended Data Figure 9 | MDCs and EDCs of the ARPES spectral cuts in Figure 4 and Extended Data Figure 10. a, b,** MDCs corresponding to the spectral cuts measured by LH and LV-polarized 103 eV photons in Fig. 4 for 2323 structure. **c**, **d**, Corresponding EDCs for 2323 structure. **e-h**, Corresponding MDCs and EDCs for 2222 structure in Extended Data Fig. 10b. **i-l**, Corresponding MDCs and EDCs for 1212 structure in Extended Data Fig. 10c. Note that the intensity scales of MDCs and EDCs have been adjusted to highlight the peaks positions.

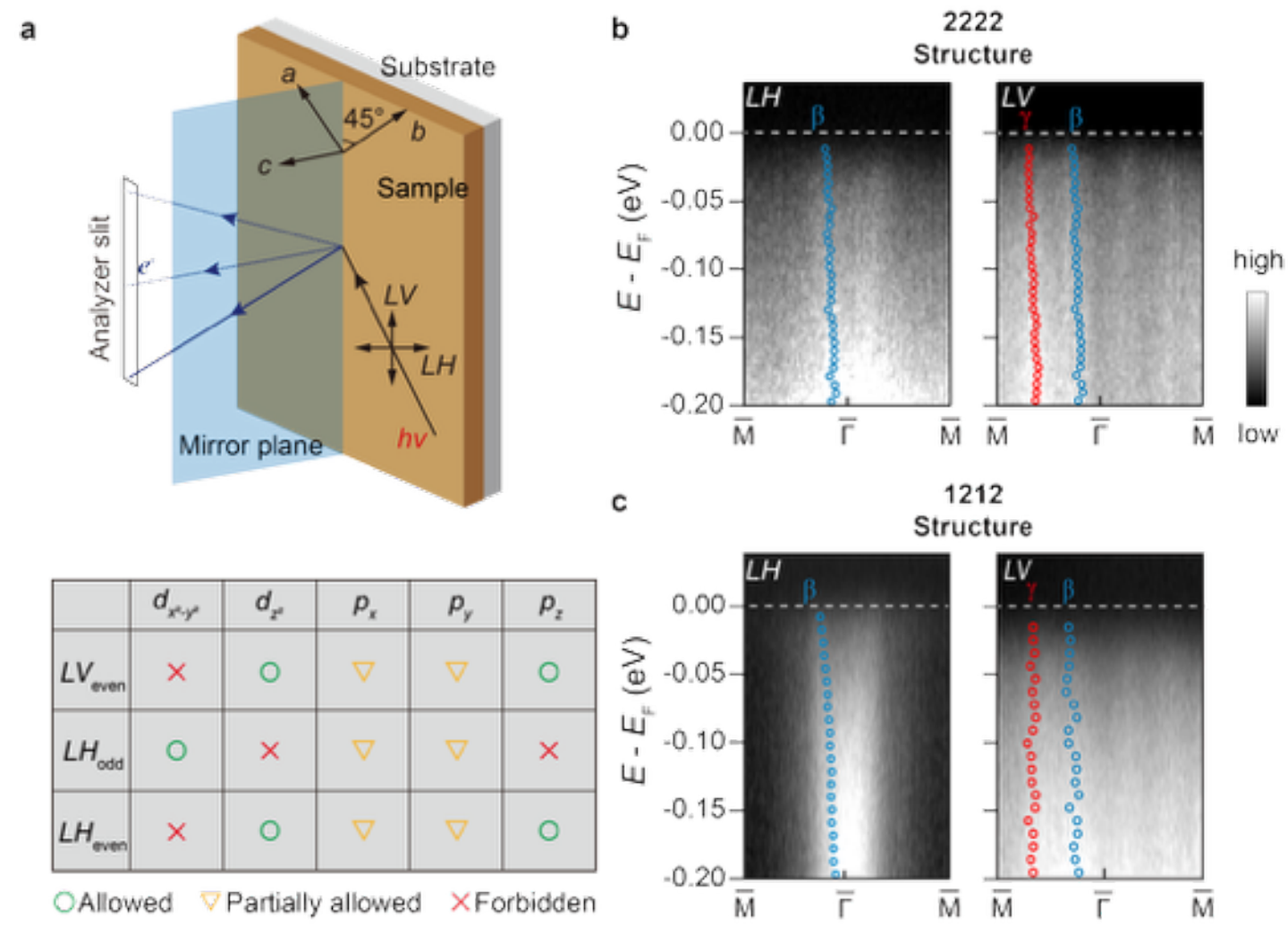

| | $d_{x^2-y^2}$ | $d_{z^2}$ | $p_x$ | $p_y$ | $p_z$ |
|---|---|---|---|---|---|
| $LV_{even}$ | × | ○ | ▽ | ▽ | ○ |
| $LH_{odd}$ | ○ | × | ▽ | ▽ | × |
| $LH_{even}$ | × | ○ | ▽ | ▽ | ○ |

**Extended Data Figure 10 | Matrix element effects and polarization dependence of the 2222 and 1212 structure. a,** The experimental geometry of the ARPES measurements with polarized photons (upper panel) and the possibility of the detection of *d* and *p* orbitals with LH and LV photons (lower panel). **b** The ARPES spectral cuts along $\overline{\mathrm{M}}$-$\overline{\Gamma}$-$\overline{\mathrm{M}}$ measured by linear horizontal (left panel) and linear vertical (right panel) 103 eV photons for 2222 structure film. **c**, Corresponding spectra cuts of 1212 superstructure film. The color bar indicates spectral intensity.